\documentclass[10pt]{article}

\usepackage{graphicx}
\usepackage{amsmath,amssymb}

\usepackage{titlesec}
\titleformat{\section}{\normalsize\bfseries}{\thesection}{1em}{}
\usepackage[font=small]{caption}
\usepackage{setspace} 

\title{\LARGE Photonuclear production of vector mesons in ultra-peripheral Pb-Pb collisions at the LHC}

\author{}

\date{}

\begin{document}

\maketitle 

\centerline{\normalsize Joakim Nystrand (for the ALICE Collaboration)}

\vspace{0.2cm}

\centerline{\small \it Department of Physics and Technology}
\centerline{\small \it University of Bergen, Bergen, Norway}
\centerline{\small \it Joakim.Nystrand@ift.uib.no}
\vspace{2.0cm}

\begin{abstract}
\small
Vector mesons are copiously produced in ultra-peripheral nucleus-nucleus collisions. In these collisions, 
the nuclei are separated by impact parameters larger than the sum of the nuclear radii, and the interaction 
is mediated by the electromagnetic field. The interaction effectively corresponds to a photonuclear 
interaction between a photon, generated from the electromagnetic field of one of the nuclei, and the target 
nucleus. The ALICE Collaboration has previously published results on exclusive $J/\psi$ photoproduction at mid and 
forward rapidities in Pb-Pb collisions. The cross section for this process is a particularly good measure of 
the nuclear gluon distribution. In this talk, the latest results on exclusive production of light and heavy 
vector mesons from ALICE in Pb-Pb collisions will be presented. \\

\vspace{0.2cm}
\end{abstract}

\begin{small}

\section{Introduction}

The strong electromagnetic fields accompanying heavy-ions accelerated at the LHC may lead to particle production 
in ultra-peripheral collisions where there is no overlap between the colliding nuclei. This has been exploited 
by the ALICE Collaboration to constrain the nuclear and proton gluon distributions through exclusive photoproduction 
of J/$\psi$ mesons~\cite{Abelev:2012ba,Abbas:2013oua,TheALICE:2014dwa}. This presentation extends these studies to 
exclusive photoproduction of $\rho^0$ and $\psi$(2S). It also contains results on two-photon production 
of $e^+e^-$ pairs in a novel invariant mass range. 

During the heavy-ion runs in 2010 and 2011, ALICE used dedicated triggers for collecting vector mesons 
produced in ultra-peripheral collisions. The ALICE detector and the triggers used during the 2011 run are described 
in~\cite{Abelev:2012ba,Abbas:2013oua}. The midrapidity triggers were based on input from the ALICE Time-of-Flight 
(TOF), Silicon Pixel (SPD), and VZERO detectors. At least two hits were required in TOF and SPD, while the VZERO 
detectors, covering $2.8 < \eta < 5.2$ and $-3.7 < \eta < -1.7$, were used as veto detectors and were required to be 
empty. In 2011, a cut on azimuthal angle was applied which required the hits in TOF 
to be back-to-back, thereby restricting the final states to have invariant masses $\geq$ 2 GeV/c$^2$. The 
ultra-peripheral trigger used in 2010 was similar to the one in 2011 but without the angular cut. During 
the early part of the 2010 run a trigger using input only from the TOF was also used. The integrated luminosities 
for the analyses presented below were about 260 mb$^{-1}$ ($\rho^0$ and $e^+e^-$ pairs, 2010 data) and 22 $\mu$b$^{-1}$ 
($\psi$(2S), 2011 data). 
\vspace{2.0cm}

\begin{figure}
\begin{center}
\includegraphics*[width=.46\textwidth]{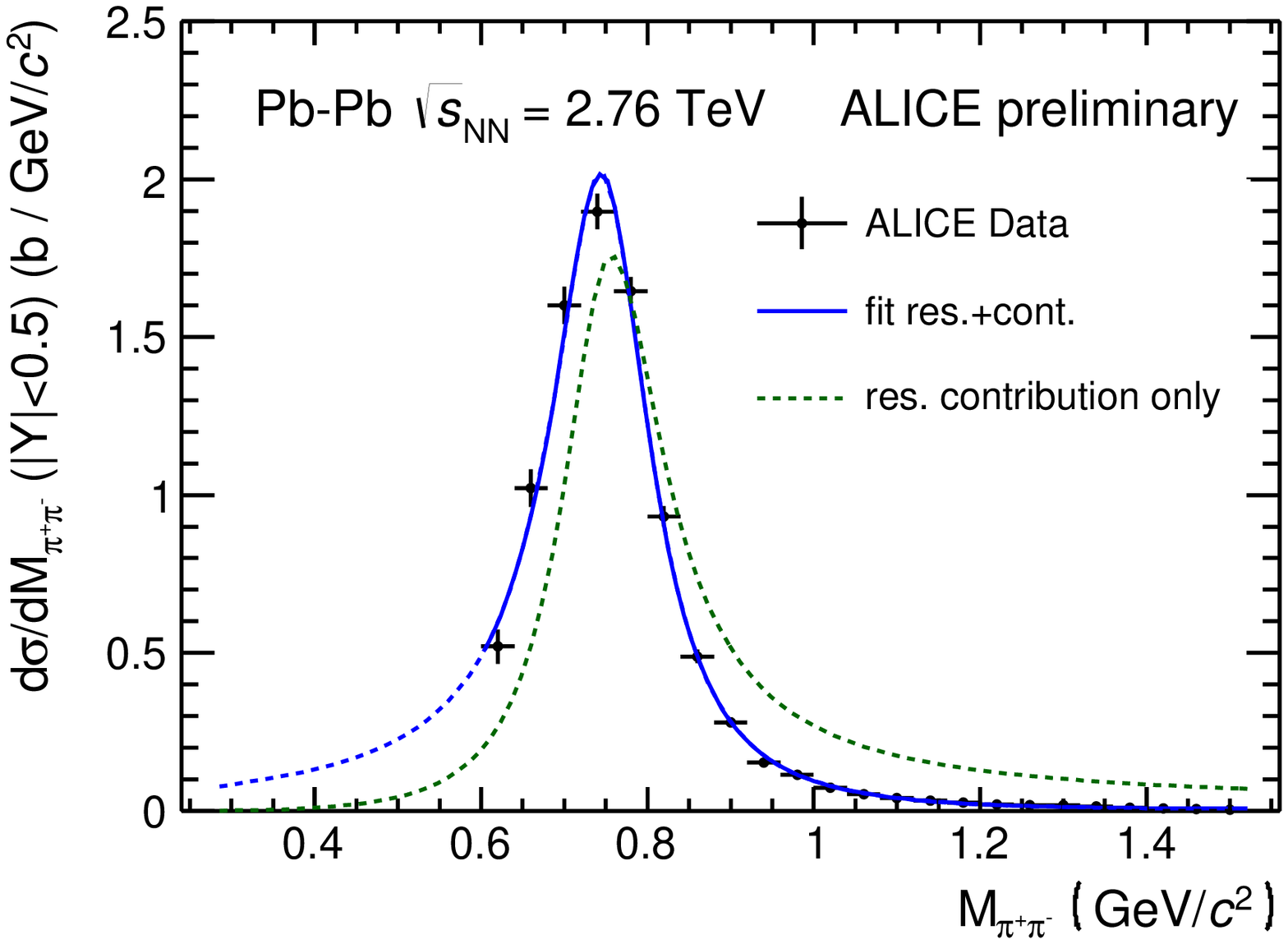}
\includegraphics*[width=.46\textwidth]{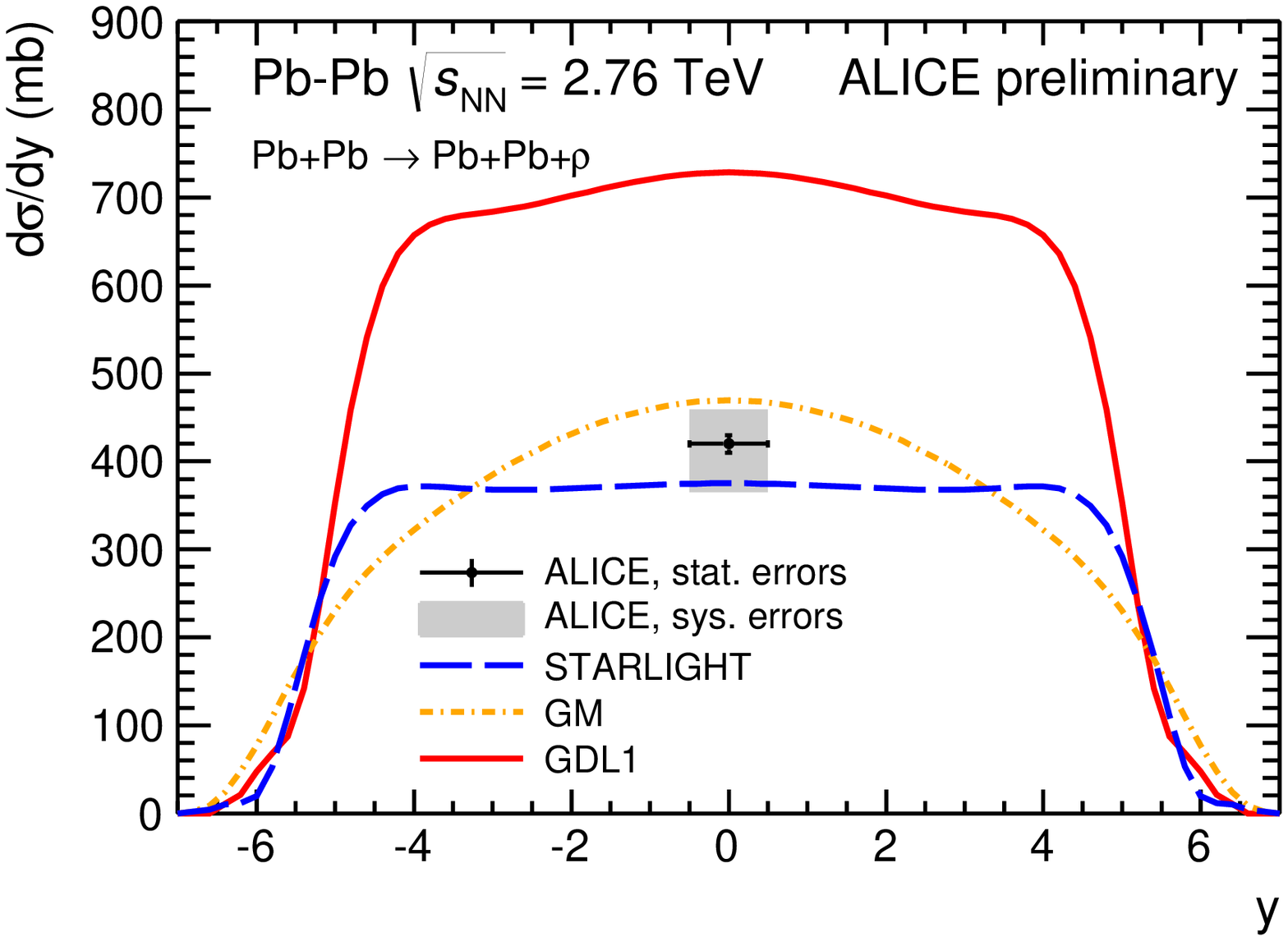}
\caption{Left: Invariant mass distribution of exclusive $\pi^+\pi^-$ pairs with $p_T <$~0.15 GeV/c. The curves
show the fit to a Breit-Wigner distribution, as discussed in the text. Right: The 
measured $\mathrm{d}\sigma/\mathrm{d}y$ for coherent photoproduction of $\rho^0$ compared with model predictions.}
\label{fig:rho}
\end{center}
\end{figure}

\section{Coherent $\rho^0$ production}

The events used for the analysis were required to satisfy the ultra-peripheral trigger, to have exactly  
two reconstructed tracks identified as pions from the dE/dx in the Time Projection Chamber (TPC), and to have 
a reconstructed primary vertex 
within $\pm$10~cm from the center of the interaction region along the beam axis. To select coherent production 
the pair-$p_T$ was required to be $< 0.15$~GeV/c. The rapidity of the pair was 
required to be within $|y|<0.5$ to avoid edge effects. The data were corrected for acceptance and efficiency bin-by-bin 
in invariant mass using simulated 
events processed through the detector response simulation and reconstructed back using the same selection as 
for real data. The simulated events consisted of two pions of opposite charge with a flat distribution in 
invariant mass ($2 m_{\pi} \leq M_{\pi\pi} \leq$~1.5~GeV/c$^2$), rapidity ($|y|<0.5$), and transverse momentum 
($p_T <$~0.15 GeV/c). 
The decay angle distribution was not measured. In the simulation, it was, as 
expected for the decay from a spin-1 state, 
assumed to be  $\mathrm{d}n/\mathrm{d}\cos(\theta) \propto \sin^2(\theta)$ in the $\pi \pi$ center of mass. 
The systematic error is estimated and 
includes the errors in signal extraction, track selection, luminosity, trigger efficiency, incoherent 
contribution, and particle identification.

The corrected invariant mass distribution is shown in Fig.~\ref{fig:rho} (left). The distribution has been fitted 
to a relativistic Breit-Wigner function plus a continuum term: 
\begin{equation}
\frac{\mathrm{d}\sigma}{\mathrm{d}M_{\pi\pi}} =
\left|
A\frac{\sqrt{M_{\pi\pi} M_{\rho} \Gamma(M_{\pi\pi})}}
{M^2_{\pi\pi} - M^2_{\rho} + i M_{\rho} \Gamma(M_{\pi\pi})}+B
\right|^2 \; \; \; \rm{with} \; \; \;
\Gamma(M_{\pi\pi}) = \Gamma_{\rho} \frac{M_{\rho}}{M_{\pi\pi}}
\left[
\frac{M^2_{\pi\pi}-4m^2_\pi}{M^2_{\rho}-4m^2_\pi}
\right]^{\frac{3}{2}} \; , 
\end{equation}
where $A$ and $B$ are the amplitudes for the resonant and continuum contributions,  respectively.
$M_{\rho}$ and $\Gamma_{\rho}$ are the mass and width of the $\rho^0$ meson, and
$m_\pi$ is the mass of the pion. Although other parameterizations of the $\rho^0$ shape are possible, the 
above is chosen to be consistent with the most recent $\rho^0$ photoproduction measurements by ZEUS~\cite{Breitweg:1997ed} 
and STAR~\cite{Abelev:2007nb}. 
The fit gives $M_{\rho} = 761.6 \pm 2.3 (stat.) ^{+6.1}_{-3.0} (sys.)$~MeV/c$^2$ and 
$\Gamma_{\rho} = 150.2 \pm5.5 (stat.) ^{+12.0}_{-5.6}(sys.)$~MeV/c$^2$, in good agreement with the PDG values. The ratio 
of the non-resonant to resonant amplitudes is $|B/A| = 0.50 \pm 0.04 (stat.) ^{+0.10}_{-0.04}(sys.)$ (GeV/c$^2$)$^{-1/2}$. 

The cross section for coherent $\rho^0$ production is obtained by integrating the resonant part of the 
$\mathrm{d}\sigma/\mathrm{d}M_{\pi\pi}$ (dashed curve in Fig.~\ref{fig:rho} (left)) from $2 m_{\pi}$ to 1.5 GeV/c$^2$. 
The contribution from incoherent 
photoproduction below $p_T <$ 0.15~GeV/c is estimated to be 5\% and is subtracted. The result is 
$\mathrm{d}\sigma/\mathrm{d}y = 420 \pm 10 (stat.) ^{+39}_{-55} (sys.)$~mb, which 
is shown and compared with model predictions in Fig.~\ref{fig:rho} (right). 

The measured cross section is in agreement with STARLIGHT~\cite{starlight} and the calculation by Goncalves 
and Machado (GM)~\cite{Goncalves:2011vf} while the GDL (Glauber-Donnachie-Landshoff) prediction~\cite{Frankfurt:2002wc} 
is about a factor of 2 higher than data. The calculation by GM is based on the Color Dipole model, while STARLIGHT and 
GDL use the photon-proton cross section $\sigma(\gamma+p \rightarrow \rho^0+p)$ constrained from data as input. All 
calculations use the Glauber model to scale the cross section from $\gamma$-nucleon to $\gamma$-nucleus. 
The agreement with STARLIGHT is a bit surprising since the calculation 
does not include the elastic part of the total cross section, which is included in the GDL model, but a similar
trend was observed by STAR at RHIC~\cite{Abelev:2007nb}. 

\section{Coherent $\psi$(2S) production}

Photoproduced $\psi$(2S) mesons can be studied in the dilepton decay channel using the ALICE triggers 
for ultra-peripheral collisions~\cite{Abbas:2013oua}. The trigger at midrapidity is, however, also sensitive 
to the decay channel $\psi(2\mathrm{S}) \rightarrow J/\psi + \pi^+ \pi^-$, followed by $J/\psi \rightarrow e^+e^-/\mu^+\mu^-$, 
which has a more advantageous branching ratio of 34\%. These events are characterized by two hard tracks with 
$p_T >$~1~GeV/c from the decay of the $J/\psi$ and two soft tracks with $p_T <$~0.4~GeV/c from the two pions. 

The triggered events were required to have exactly 2 or 4 reconstructed tracks and a reconstructed primary 
vertex. The dileptons can be separated into $e^+e^-$ or $\mu^+\mu^-$ using the TPC dE/dx. To select coherently 
produced $\psi$(2S) a requirement $p_T <$~0.15 (0.30) GeV/c was applied for the decay 
channels containing muons (electrons). 

The yield in the direct dilepton channel ($e^+e^-$ and $\mu^+\mu^-$ combined) was obtained by fitting the 
invariant mass distribution to the sum of a Crystal Ball function for the signal and an exponential for the 
$\gamma \gamma \rightarrow l^+l^-$ background. 
The 4 track samples have a very high purity and the yield is obtained by counting the number of events 
within selected intervals in phase space determined from Monte Carlo simulations. The invariant mass of 
the dileptons are required to be within $3.0 \leq M_{inv} \leq 3.2$ ($2.6 \leq M_{inv} \leq 3.2$) GeV/c$^2$ 
for $\mu^+\mu^-$ ($e^+e^-$) to select pairs coming from a $J/\psi$ decay. The $\mu^+\mu^-\pi^+\pi^-$ 
($e^+e^-\pi^+\pi^-$) final state is required to have an invariant mass $3.6 \leq M_{inv} \leq 3.8$ 
($3.1 \leq M_{inv} \leq 3.8$) GeV/c$^2$ to count as a $\psi$(2S). This gives a total of 17 (11) events 
for the $\mu^+\mu^-\pi^+\pi^-$ ($e^+e^-\pi^+\pi^-$) decay channel, with a like-sign background of 1 (0) 
event.

The extracted yield is corrected for acceptance and efficiency using STARLIGHT events processed 
through the detector response simulation and reconstructed back using the same selection as for real data. 
The cross section is obtained from a weighted mean of the dilepton and 4--track decay channels, giving 
$\mathrm{d}\sigma/\mathrm{d}y = 0.83 \pm 0.19 (stat.+sys.)$~mb. This is compared to model 
predictions~\cite{starlight,Adeluyi:2013tuu,Ducati:2013bya,Lappi:2013am} in Fig.~\ref{fig:psi2Sxsection}. 

\begin{figure}
\begin{center}
\includegraphics*[width=.67\textwidth]{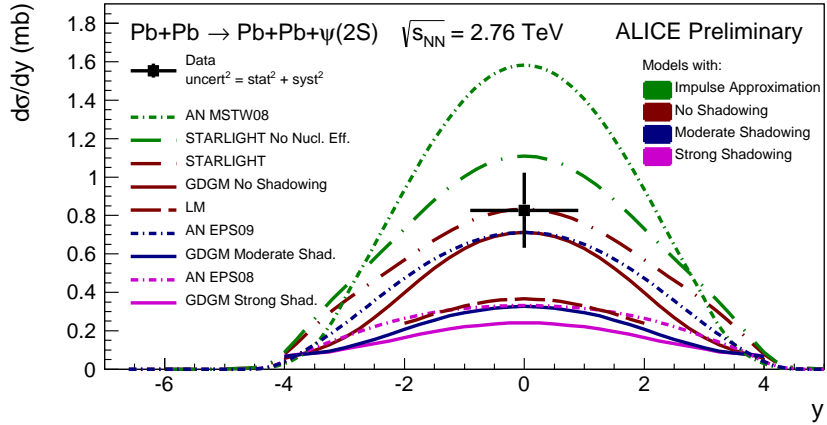}
\caption{Measured $\mathrm{d}\sigma/\mathrm{d}y$ for coherent $\psi$(2S) photoproduction compared with model calculations 
(STARLIGHT~\cite{starlight}, AN~\cite{Adeluyi:2013tuu}, GDGM~\cite{Ducati:2013bya}, LM~\cite{Lappi:2013am}).
}
\label{fig:psi2Sxsection}
\end{center}
\end{figure}

The $\psi$(2S) is a hard probe, the scale being set by its mass, and is therefore expected to be sensitive to 
the nuclear gluon distribution in the same way as the $J/\psi$. The experimental error is, however, larger 
for the $\psi$(2S) than for the $J/\psi$ measurement, mainly because of the limited statistics. It is 
also important to note that the underlying $\gamma+p \rightarrow V+p$ cross section has a considerably larger 
uncertainty for $\psi$(2S) than for $J/\psi$. This is illustrated by the large discrepancy between the 
AN~\cite{Adeluyi:2013tuu} and STARLIGHT predictions without nuclear effects in the figure. 
One can nevertheless conclude that 
the measured cross section disfavors models with no nuclear effects and models with strong gluon shadowing. 

\section{Two-photon production of $e^+e^-$ pairs}

Two-photon production of $e^+e^-$ pairs has a topology similar to that of exclusive vector mesons production 
followed by decay into a pair of dileptons or pions. This process is of interest since the coupling between the 
photon and the emitting nucleus is enhanced by a factor Z (the nuclear charge). One can therefore expect higher 
order terms to be important. The results from ALICE on $\gamma \gamma \rightarrow e^+e^-$ from the 2011 Pb-Pb run were, 
however, found to be in good agreement with STARLIGHT, which includes only the leading order QED 
terms~\cite{Abbas:2013oua}. Because of the requirement on azimuthal angle in the trigger, the results from 2011 
were restricted 
to high invariant masses, $M_{ee} >$~2.2~GeV/c$^2$. The 2010 data allow to go down to $M_{ee} =$~0.6~GeV/c$^2$. 

\begin{figure}
\begin{center}
\includegraphics*[height=4.9cm]{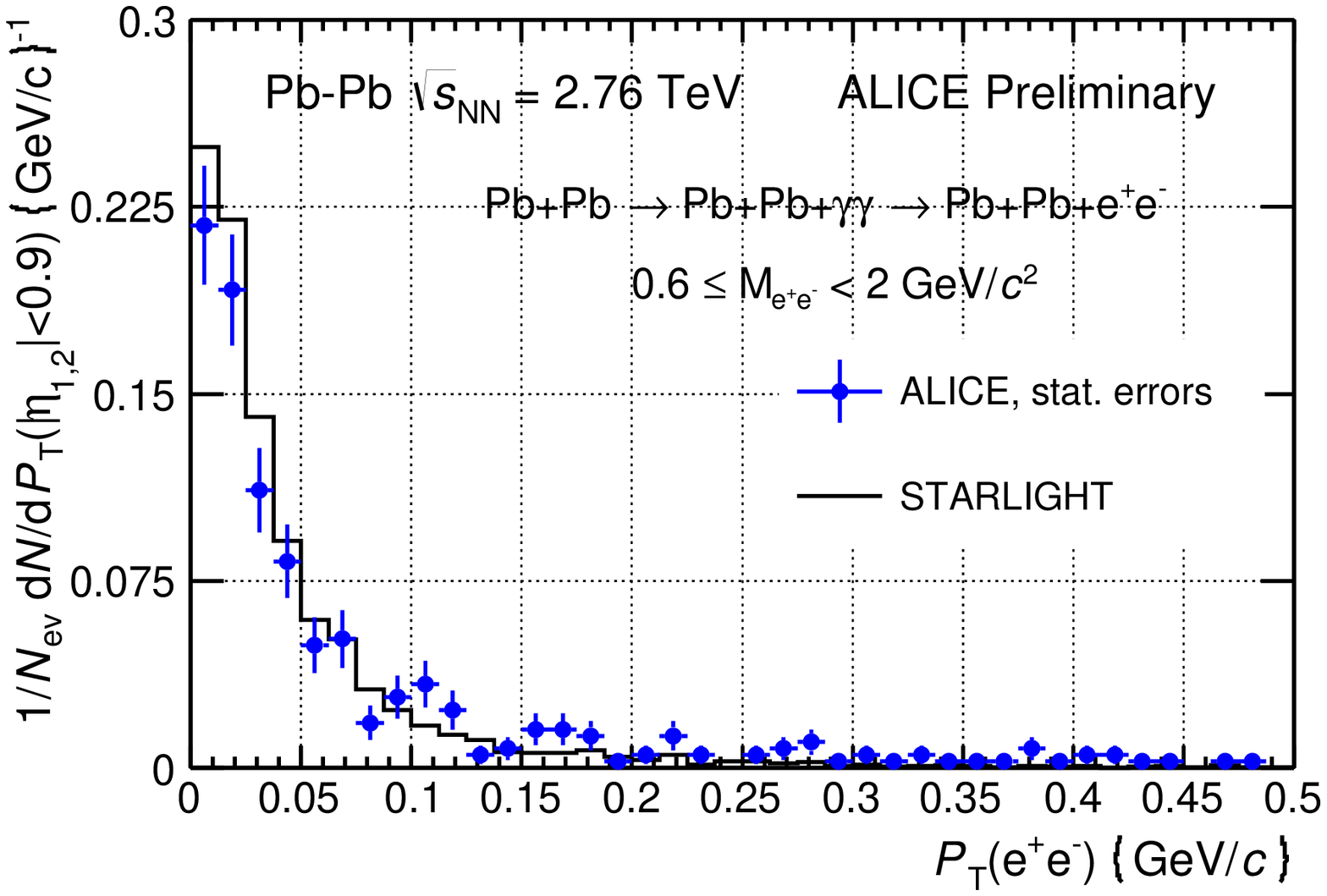}
\includegraphics*[height=4.9cm]{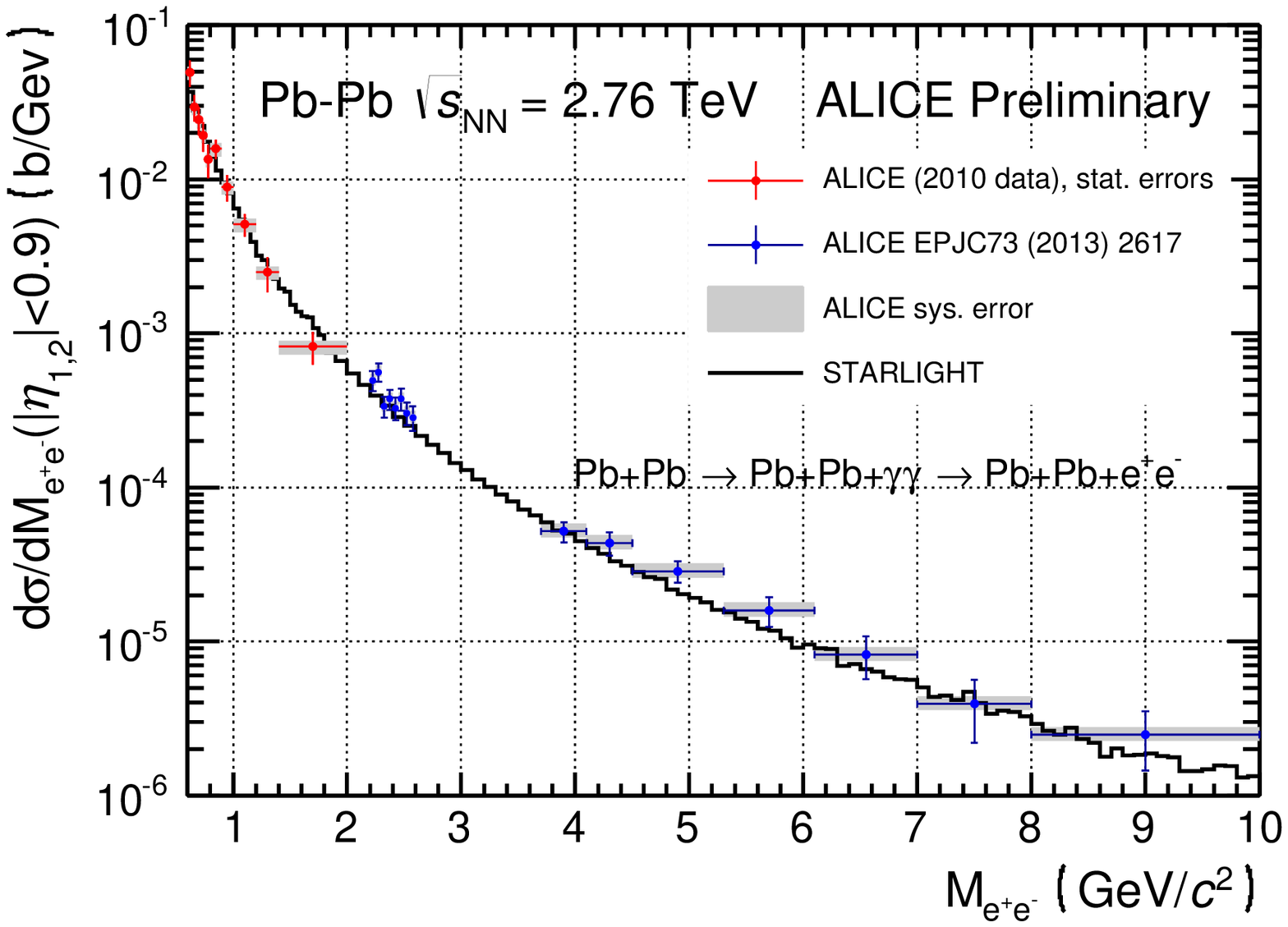}
\caption{Left: Transverse momentum distribution of exclusive $e^+e^-$ pairs with $0.6 \leq M_{ee} \leq 2.0$~GeV/c$^2$ 
compared with STARLIGHT. Right: Differential cross section, $\mathrm{d}\sigma/\mathrm{d}M_{ee}$, for 
$Pb+Pb \rightarrow Pb+Pb+e^+e^-$ compared with STARLIGHT.}
\label{fig:gg}
\end{center}
\end{figure}

The event selection was similar to that for coherent $\rho^0$ production, with the selection on the TPC dE/dx 
modified to accept electrons rather than pions. The raw data were corrected for acceptance and efficiency using 
events generated by STARLIGHT. Figure~\ref{fig:gg} (left) shows the $p_T$ distribution of the selected 
$e^+e^-$ pairs. The distribution is well described by STARLIGHT, indicating that there is no background not 
accounted for in the sample. The measured cross section for the selection $0.6 \leq M_{ee} \leq 2.0$~GeV/c$^2$ 
and $|\eta_{1,2}| <$~0.9 ($\eta_{1,2}$ are the pseudorapidities of the two tracks) is 
$9.8 \pm 0.6 (stat.) ^{+0.9}_{-1.2} (sys.)$~mb. The STARLIGHT prediction for the same selection is $\sigma =$~9.7~mb. 
The differential cross section, $\mathrm{d}\sigma/\mathrm{d}M_{ee}$, is shown in Fig.~\ref{fig:gg} (right) 
together with the previous ALICE measurement for $M_{ee} >$~2.2~GeV/c$^2$ and the cross section from STARLIGHT. 
The variation of $\mathrm{d}\sigma/\mathrm{d}M_{ee}$ with $M_{ee}$ is well described by STARLIGHT.

\section{Conclusions}

The ALICE Collaboration has made the first measurements of coherent $\rho^0$ and $\psi$(2S) photoproduction 
in Pb-Pb collisions at the LHC. The measured cross section for the $\psi$(2S) disfavors models with no nuclear 
effects and models with strong gluon shadowing. This is thus consistent with the conclusions from the $J/\psi$ 
measurements~\cite{Abelev:2012ba,Abbas:2013oua} but less strong because of the larger uncertainties for $\psi$(2S) 
discussed above. The results on $\rho^0$ show that a straightforward scaling of the $\gamma$-p cross section 
using the Glauber model~\cite{Frankfurt:2002wc} overpredicts the cross section. This confirms what was observed in 
Au-Au collisions at RHIC energies. The cross section for two-photon production of $e^+e^-$ pairs at midrapidity 
in Pb-Pb collisions is in good agreement with leading order QED as implemented in STARLIGHT in the invariant 
mass range $0.6 \leq M_{ee} \leq 10.0$~GeV/c$^2$.

\end{small}

\end{document}